\documentclass[reprint, aps, nofootinbib]{revtex4-2}
\usepackage{graphicx}                            
\usepackage{amsmath} 
\usepackage{amssymb}

\usepackage{braket}
\usepackage[colorlinks=true,bookmarks=false,citecolor=blue,urlcolor=blue]{hyperref}
\usepackage{float}

\newcommand{\Rsp}{R_{\text{sp}}}

\newcommand{\Lambdath}{\Lambda_{\text{th}}}

\begin{document}

\title{Intense squeezed light from lasers with sharply nonlinear gain at optical frequencies}
\author{Linh Nguyen$^{1}$, Jamison Sloan$^{2}$, Nicholas Rivera$^{1,3}$, Marin Solja\v{c}i\'{c}$^{1}$}

\address{$^1$ Department of Physics, Massachusetts Institute of Technology, Cambridge MA, United States. \\
$^2$Department of Electrical Engineering and Computer Science, Massachusetts Institute of Technology, Cambridge MA, United States. \\
$^3$ Department of Physics, Harvard University, Cambridge MA, United States.}
\email{linhnk@mit.edu}

\date{\today} 

\begin{abstract}
\noindent	

Non-classical states of light, such as number-squeezed light, with fluctuations below the classical shot noise level, have important uses in metrology, communication, quantum information processing, and quantum simulation. However, generating these non-classical states of light, especially with high intensity and high degree of squeezing, is challenging. To address this problem, we introduce a new concept which uses gain to generate intense sub-Poissonian light at optical frequencies. It exploits a strongly nonlinear gain for photons which arises from a combination of frequency-dependent gain and Kerr nonlinearity. In this laser architecture, the interaction between the gain medium and Kerr nonlinearity suppresses the spontaneous emission at high photon number states, leading to a strong ``negative feedback'' that suppresses photon-number fluctuations. We discuss realistic implementations of this concept based on the use of solid-state gain media in laser cavities with Kerr nonlinear materials, showing how 90\% squeezing of photon number fluctuations below the shot noise level can be realized.

\end{abstract}

\maketitle

\pagestyle{myheadings}
\markboth{Linh Nguyen}{Intense squeezed light from lasers with sharply nonlinear gain}
\thispagestyle{empty}

Non-classical states of light such as squeezed states and Fock states are famous for maintaining noise levels below the classical ``shot noise'' limit. Therefore, these states are considered important for applications in precision measurement \cite{davidovich1996sub, thomas2011real}, optical communication \cite{teich1989squeezed}, quantum information processing, and quantum simulation \cite{aaronson2011computational,lund2014boson,huh2015boson,wang2017high, hamilton2017gaussian, brod2019photonic, wang2020efficient}. Intense coherent light is readily produced by lasers and masers. However, generation of squeezed light necessarily relies on some types of nonlinear mechanism, which is often form of nonlinear loss or gain.

In the microwave regime, various mechanisms of nonlinear gain have been explored as tools to modify the quantum statistics of light. For example, experiments with so-called ``micromasers'' pumped by excited Rydberg atoms have been used to generate sub-Poissonian light in a microwave resonator \cite{rempe1990observation, varcoe2000preparing, sayrin2011real,uria2020deterministic,canela2020bright}. Similar physics can also be exploited by superconducting qubits to deterministically generate Fock states up to order 10 \cite{hofheinz2008generation,heeres2017implementing}. Robust effects like photon blockade have also been used as sources of single photons \cite{birnbaum2005photon} and sub-poissonian light \cite{flayac2017unconventional}. Additionally, it has been shown how light-matter systems in the so-called ``deep strong coupling'' regime can be used to create an effective $N$-photon blockade which lends itself to the generation of $N$-photon Fock states when used as a maser gain medium \cite{Rivera2021Fock}. 

Despite the effectiveness of the described tools in the microwave regime, many of these techniques do not readily extend to optical frequencies. In the optical regime, both second and third-order nonlinear interactions, as well as nonlinear dissipation mechanisms such as two-photon absorption have been used to squeeze light below the shot noise limit \cite{two_photon_absoprtion, ritsch1990quantum, ritsch1992quantum,kitagawa1986number, shirasaki1990squeezing,bergman1991squeezing,kitching1994amplitude,schmitt1998photon,andersen201630}, with many promising new approaches for very strong reduction of intensity noise, especially for lower intensity light \cite{gonzalez2017efficient,kilin1995fock,leonski1997fock, munoz2014emitters,yanagimoto2019adiabatic,ben2021shaping,lingenfelter2021unconditional,rivera2023creating}. However, these methods have typically been limited to the generation of sub-Poissonian light which is only 50\% below the shot noise limit, with notable exceptions limited to specialized cases (e.g., requiring the use of solitons in fibers) \cite{andersen201630}. Furthermore, by and large,  many of these nonlinear techniques work either at very low powers (e.g., below the threshold of an optical parametric oscillator), or moderate powers. It is still an open question of how to realize strongly squeezed light at macroscopic intensities (e.g., watt-scale and beyond). In general, high power optical sources are produced by amplification through gain media, which either amplify fluctuations, or at best produce approximations to coherent states (but often have substantial excess noise). As a result, the generation of intense, and strongly sub-Poissonian optical radiation remains a great challenge.

Here, we introduce a mechanism of sharply nonlinear gain which can be used to create intense sub-Poissonian light at optical frequencies. We show how this nonlinear gain can be realized by incorporating Kerr nonlinearity into a traditional solid state laser architecture. We detail how the nonlinear gain acts as a form of ``super-saturation,'' leading to new phenomena like nonlinear power curves, and the suppression of transient relaxation oscillations. Furthermore, we show how this nonlinear gain can lead to macroscopic states of light with intensity fluctuations 90\% below the shot noise level, which corresponds to a sub-Poissonian (number-squeezed) states of light that have no classical analog \cite{mandel1995optical, walls2007quantum}. This is in strong contrast to conventional high-power lasers, which produce coherent states or states with substantial excess intensity noise.

\section{A laser based on Kerr nonlinearity}

\begin{figure}[t]
    \includegraphics[width=\linewidth]{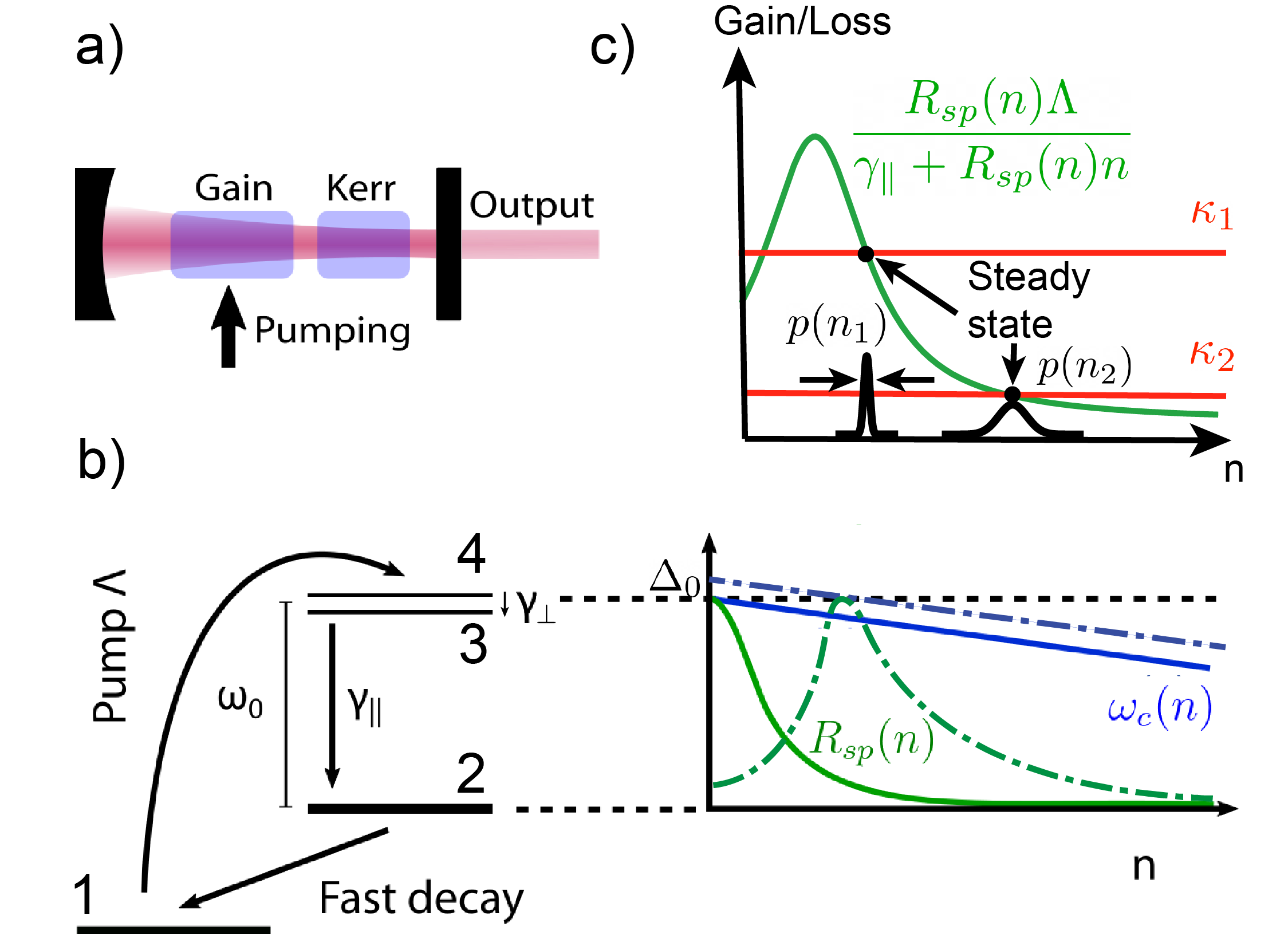}
    \caption{\textbf{A laser based on Kerr nonlinearity.} (a) A pumped gain medium coupled to a cavity with Kerr nonlinearity. (b) Four-level atoms in the gain medium, with lasing transition between level 3 and 2 with frequency $\omega_0$ . 
    The dashed lines show the dependence of the cavity resonance frequency $\omega_c(n)$ (blue) and the spontaneous emission rate $\Rsp$ (green) on the light intensity of an ``off-resonant Kerr laser" ($\omega_c - \omega_0 = \Delta_0$), and the solid lines are those of ``resonant Kerr laser" ($\omega_c = \omega_0$). 
    (c) The photon number probability distribution (black) depends on the angle between the gain (green) and the loss (red) at their intersection point (or the steady state). Because the gain intersects with the loss $\kappa_1$ at intensity $n_1$ more steeply than with the loss $\kappa_2$ at intensity $n_2$, the probability distribution at $n_1$ is more squeezed, thus the photon number variance is reduced.}
    \label{fig:my_label}
\end{figure}

The concept for the ``Kerr laser" is illustrated in Fig. 1a. The system consists of a pumped gain medium in a cavity containing a Kerr nonlinear medium. The Hamiltonian of just a cavity with an embedded Kerr medium (so no gain) is $H_{\text{Kerr}}/\hbar = \omega_c \hat{a}^\dagger \hat{a} - \frac{\beta}{2} \hat{a}^\dagger \hat{a} - \frac{\beta}{2} \hat{a}^{\dagger 2} \hat{a}^2$ ($\hat{a}$ and $\hat{a}^\dagger$ are the annihilation and creation operators of the cavity mode, $\beta$ is the Kerr nonlinear strength of a single photon, and $\omega_c$ is the cavity frequency) \cite{walls2007quantum, drummond1980quantum}. In such a system, the energy needed to add one more photon into a cavity which already contains $n - 1$ photons is $\hbar(\omega_n - \omega_{n-1}) = \hbar \omega_c(n) = \hbar(\omega_c - \beta n)$. The cavity resonance frequency depends linearly on the photon number as shown in Fig. 1b. Hence, at a certain Fock state $|n\rangle$ in the cavity, the gain medium and the cavity are resonant (Fig. 1b)
Therefore, the spontaneous emission rate $\Rsp$ is maximal for photon numbers where the cavity and the gain are resonant, but suppressed for photon numbers where they are detuned. As a result, the cavity ``prefers'' photon numbers which are close to resonance, and suppresses photon numbers away from them, leading to a reduction in the photon number fluctuations, and thus squeezing \cite{davidovich1996sub}. This type of gain can be thought of as a ``super-saturable" gain, which decreases more sharply with $n$ from the equilibrium point (of balanced gain and loss), as compared to conventional saturable gain \cite{Rivera2021Fock}. Suprisingly, this approach can lead to very large photon-number squeezing in the cavity (approaching 10 dB), as we will now show.

For concreteness, we consider the gain medium to consist of typical four-level atoms (with fast non-radiative decay to render it as an approximate two-level system), such as Nd:YAG. In such systems, the polarization decay rate $\gamma_\perp$ is much larger than inversion decay rate $\gamma_\parallel$ and cavity leakage rate $\kappa$. Hence, we can adiabatically eliminate the polarization degree of freedom, resulting in two Heisenberg-Langevin equations of motion for the atomic inversion $S$ and the light intensity $n$: 
\begin{subequations}
\begin{align}
    \Dot{n} &= \left(\Rsp(n) S - \kappa\right)n + F_n\\
    \Dot{S} &= \Lambda - \left(\gamma_\parallel + \Rsp(n) n\right) S + F_S .
\end{align}
\label{eq:photon_num_and_inv}
\end{subequations}
The spontaneous emission rate $\Rsp(n) \equiv 2g^2 \gamma_\perp/(\gamma_\perp^2 + \Delta(n)^2)$, where $\Delta(n)$ is the detuning between the atomic frequency and the (number-dependent) resonance cavity frequency, $g$ is the coupling strength between the lasing transition and the cavity mode, and $\Lambda$ is the pump rate of the gain medium. Quantum fluctuations are incorporated into the rate equations through the Langevin forces $F_n$ and $F_S$ whose correlation functions are: $\braket{F_\mu(t) F_\nu(t')} = \braket{2D_{\mu\nu}}\delta(t- t')$, with diffusion coefficients given by $\braket{2D_{nn}} = \braket{\Rsp(n)Sn + \kappa n}$, $\braket{2D_{SS}} = \braket{\Lambda + \gamma_\parallel S + \Rsp(n) n S}$, and $\braket{2D_{Sn}} = \braket{2D_{nS}} = - \braket{\Rsp(n) n S}$.

\begin{figure}[h]
    \includegraphics[width=\linewidth]{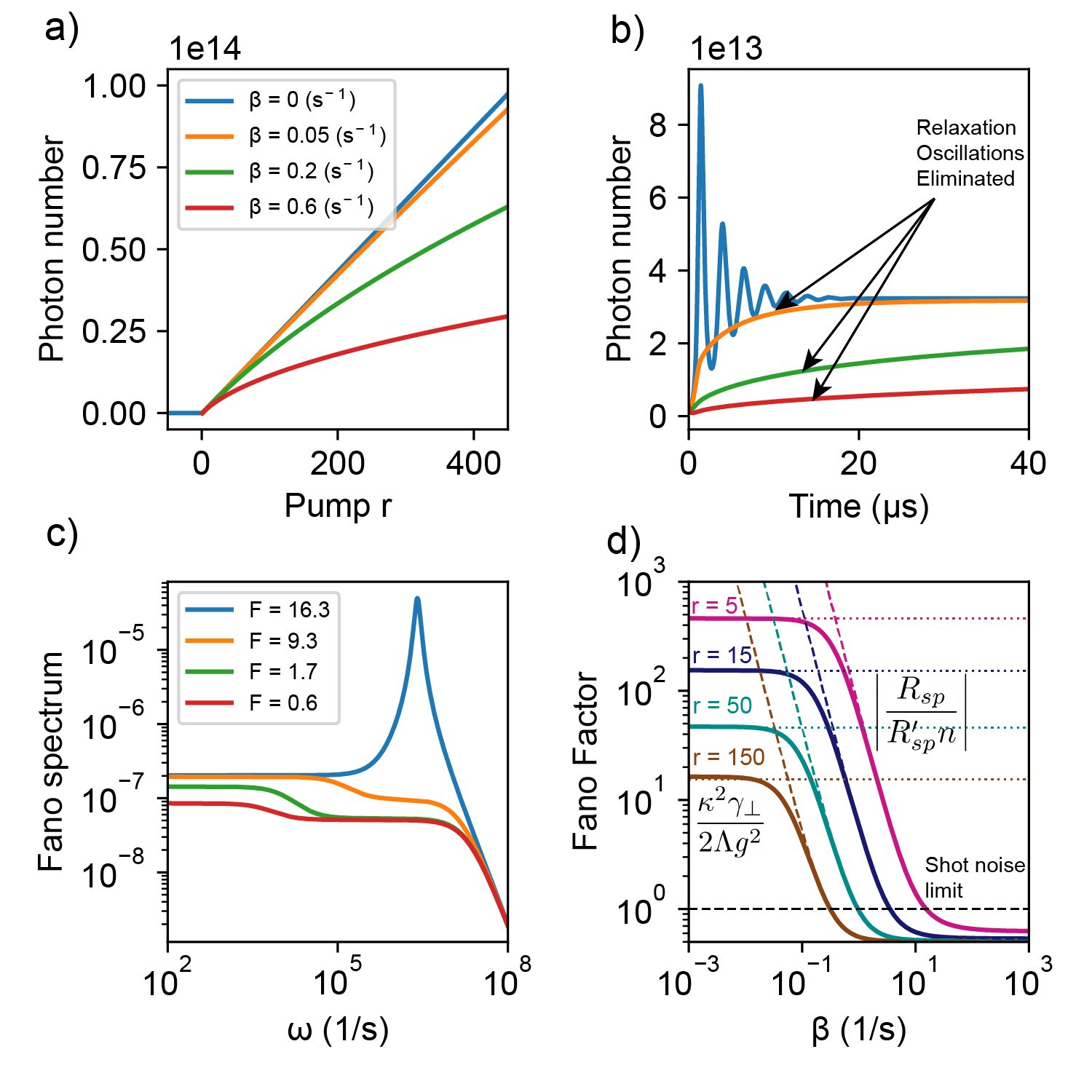}
    \caption{\textbf{Output characteristics of a ``resonant Kerr laser.''} Parameters used for an Nd:YAG based system are: $\gamma_\perp = 10^{12} s^{-1}$, $\gamma_\parallel = 4.34\times 10^3 s^{-1}$, $\kappa = 10^{7} s^{-1} $, and the coupling coefficient $g = 10^2 s^{-1}$. For figures (a), (b), and (c) lines with the same color have the same $\beta$ values. (a) Steady-state photon number of a Kerr laser as a function of relative pump rate $r$ for different nonlinear coefficients. 
    (b) Time evolution of photon number at different nonlinear coefficients and at fixed relative pump rate $r = 150$. 
    (c) The Fano spectrum (noise spectrum in frequency domain ($\omega$)/  intensity) at different nonlinear coefficients and at fixed relative pump rate $r = 150$ and their corresponding Fano factors F. (d) Fano factor $F$ at different pump rates as a function of nonlinear coefficient. At small $\beta$ value, $F$ can be approximated by $\frac{\kappa^2 \gamma_\perp}{2 \Lambda g^2}$ (dotted line). For large enough $\beta$, $F$ approximately equals to $\left| \frac{\Rsp(n_s)}{\Rsp'(n_s)n_s}\right|$  (dashed line). As $\beta$ becomes much larger than $\sqrt{\frac{\kappa g^2 \gamma_\perp}{2 \Lambda \gamma_{||}}}$,  $F$ approaches the sub-Poissonian value of 0.5.}
    \label{fig:my_label}
\end{figure}

First, we consider a ``resonant Kerr laser," where the empty cavity and the gain are in resonance ($\Delta(n) = \beta n$). When the pump rate exceeds the threshold value $\Lambdath = \kappa \gamma_\parallel \gamma_\perp/(2 g^2)$, the steady state cavity photon number $n_s$ becomes nonzero, and depends on the relative pump rate $r = \Lambda/ \Lambdath$ as shown in Fig. 2a. Compared to the ``linear laser" ($\beta = 0$), the steady-state intensity of the Kerr laser increases less rapidly with pump rate. This behavior is expected because as the intensity increases, the cavity frequency detunes from the atomic resonance, causing a sharper reduction in stimulated emission via $\Rsp(n)$ than saturable gain. The effect becomes significant when the Kerr nonlinear shift frequency at steady state $\beta n_s$ is comparable to the gain bandwidth $\gamma_\perp$. 

The nonlinearity affects fluctuations from the steady state even more drastically. Small fluctuations of the intensity $\delta n$ and the inversion $\delta S$ are governed by the Heisenberg equations:
\begin{equation}
    \begin{pmatrix} \delta \Dot{n} \\ \delta \Dot{S} \end{pmatrix} = \begin{pmatrix}
    n_s \kappa \frac{\Rsp'(n_s)}{\Rsp(n_s)} &  &\Rsp(n_s) n_s \\
    - n_s \kappa \frac{\Rsp'(n_s)}{\Rsp(n_s)} - \kappa & &  -\frac{\Lambda \Rsp(n_s)}{\kappa}
    \end{pmatrix}
    \begin{pmatrix}
    \delta n \\
    \delta S
    \end{pmatrix} + 
    \begin{pmatrix}
    F_n \\
    F_S
    \end{pmatrix}
    \label{eq:fluctuations_time}
\end{equation}
where $\Rsp'(n_s)$ is the first derivative of $\Rsp$ evaluated at the mean photon number $n_s$. 
Transient behavior can be described by taking expectation values of this system, which is equivalent to removing the zero-mean Langevin force terms. It is long-known that most solid-state and semiconductor lasers, when perturbed from equilibrium (e.g., by gain modulation, mechanical fluctuations, etc.), exhibits decaying characteristic intensity and inversion oscillations \cite{siegman1986lasers}, which are referred to as ``relaxation oscillations,'' and are considered undesirable in many applications. The ``Kerr'' laser resists fluctuations from the mean, so these relaxation oscillations are suppressed (Fig. 2b). Even a strength of nonlinearity which has a minimal effect on the intensity (i.e., $\beta = 0.05$ s$^{-1}$) leads to a complete suppression of relaxation oscillations. \footnote{The fractional change in the index of refraction in this case is on the order of 1\%, which while large for continuous-wave excitation, is feasible in pulsed settings, and we expect the treatment here (which is continuous-wave) to extend to that case}.

In addition to changing the mean field properties of the laser, the Kerr nonlinearity also influences the quantum statistics of the laser steady state. By solving Eq. \ref{eq:fluctuations_time} in the frequency domain $\omega$, we obtain the frequency dependent spectrum of the intensity noise:
\begin{equation}
    \braket{\delta n^\dagger (\omega) \delta n(\omega)} = 
    \frac{\left(\omega^2 + \Gamma_0^2\right) 2\kappa n_s}{\left(\Omega^2 + \Gamma^2/4 - \omega^2\right)^2 + \Gamma^2 \omega^2}.
    \label{eq:noise_spectrum}
\end{equation}
The relaxation oscillation frequency $\Omega$ is defined by $\Omega^2 \equiv \Rsp(n_s) n_s \kappa \left(n_s \frac{\Rsp'(n_s)}{\Rsp(n_s)} + 1 \right) - \frac{1}{4} \left(\frac{\Lambda \Rsp(n_s)}{\kappa} + n_s \kappa \frac{\Rsp'(n_s)}{\Rsp(n_s)}\right)^2$, and $\Gamma \equiv \Gamma_0 - \frac{\Rsp'(n)}{\Rsp(n)} n \kappa$ is the relaxation oscillation decay rate, where $\Gamma_0 \equiv \Lambda \Rsp(n_s)/\kappa$ is the decay rate without nonlinearity. 
This noise spectrum represents the intensity fluctuations associated with individual Fourier components of the laser field which fluctuates about its steady state. Such fluctuations may be studied with a photodiode and electronic spectrum analyzer. For the linear resonator, the relative intensity noise spectrum sharply peaks around $\Omega$, indicating relaxation oscillations driven by quantum noise. As the nonlinear strength increases, this oscillation peak is suppressed, and eventually eliminated, consistent with the transient behaviors shown for the same parameters (Fig. 2c). 

The steady-state photon number variance $(\Delta n)^2$ can be obtained by integrating the noise spectrum:
$(\Delta n)^2 = \frac{1}{\pi}\int_0^\infty d\omega \braket{\delta n^\dagger (\omega) \delta n(\omega)}$. The nonlinear gain leads to sizable reductions in the intensity noise, which we characterize by the Fano factor $F = (\Delta n)^2/n_s$ ($F = 1$ for Poissonian light). When $\beta n_s$ is comparable to $\gamma_\perp$, $F$ is suppressed by two orders of magnitude compared to that of a linear laser, and less than 1 (Fig. 2c). 

Analytically, $F = \frac{\kappa}{\Gamma}\left(\frac{\Gamma_0^2}{\Gamma^2/4 + \Omega^2} + 1\right)$, which approximately equals to $\kappa/\Gamma$ over a large range of pump rates ($\Lambdath < \Lambda < 1000\Lambdath$). The Kerr nonlinearity only affects $F$ significantly when $\beta n_s$ is comparable to $\gamma_\perp$. Under that condition,
\begin{equation}
    F \approx \left| \frac{\Rsp(n_s)}{\Rsp'(n_s)n_s} \right|.
    \label{eq:fano}
\end{equation}
$F$ can thus be decreased by making both $n_s$  and $|\Rsp'(n_s)/\Rsp(n_s)|$ as large as possible. Physically, this corresponds to situations in which $\Rsp$ (and thus gain) decreases as sharply as possible with increasing photon number. In the case of ``resonant Kerr laser," as $\beta$ increases, $\Rsp$ will fall more sharply with higher intensity, but the sharp falling gain will intersect the loss at lower intensity. This behavior causes $F$ to saturate at a sub-Poissionian value of $0.5$. 

\section{Further noise suppression with detuning between the gain medium and zero-photon cavity}

As we have discussed above, the minimum value of the Fano factor is dictated by both the steepness of $\Rsp$ and the mean photon number. To further suppress the intensity noise, the steep region of $\Rsp$ should ideally occur at the highest photon number possible. This effect can be realized by setting the $n=0$ cavity resonance frequency to be higher than the atomic frequency (off-resonant Kerr laser). Because of the Kerr effect, the detuning value decreases with light intensity, so the cavity and the gain medium become resonant at photon number $n_\text{res} = (\omega_c - \omega_0)/\beta$. At $n_\text{res}$, the spontaneous emission rate is also maximal (Fig. 1b).
The higher the detuning value $\Delta_0$, the more $\Rsp$ is shifted. We set $\Delta_0$ to be of 1\% of the gain frequency (about ten times the gain bandwidth, which of course leads to a much larger threshold). The cavity can be further detuned from the gain at the expense of higher pump rate to ensure that the gain of the empty cavity exceeds the cavity leakage rate. 

\begin{figure}[h]
    \includegraphics[width=\linewidth]{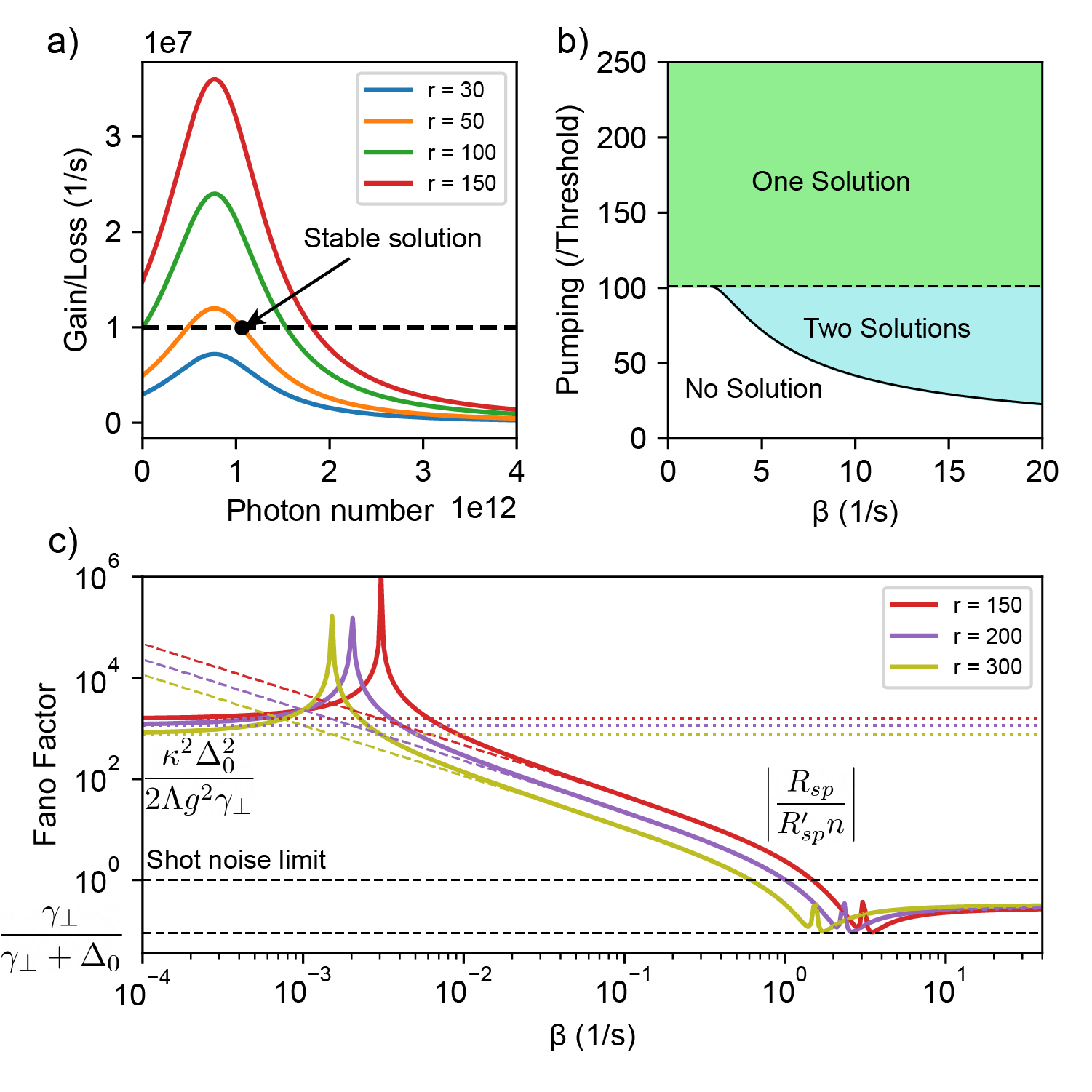}
    \caption{\textbf{Off resonant nonlinear ``Kerr laser'' with Nd:YAG gain medium and the detuning value between the gain medium and the zero photon cavity be $\Delta_0 = 10^{13} s^{-1}$.} (a) The gain and loss rate as functions of light intensity at different pump rates and at fixed nonlinear coefficient $\beta = 10 ~s^{-1}$. 
    (b) The number of steady solutions depends on both the nonlinear coefficient and pump rate. (c) Fano factor $F$ at different pump rates as a function of nonlinear coefficient. $F$ reaches a global minimum value of $\frac{\gamma_\perp}{\gamma_\perp+\Delta_0}$ when the detuning value between the cavity and the gain medium $-\Delta_0 + \beta n_s$ equals the gain bandwidth $\gamma_\perp$. At small $\beta$ value, $F \approx \kappa^2 \Delta_0^2/(2\Lambda g^2 \gamma_\perp)$ (dotted line). For large enough $\beta$ value, $F \approx \left| \frac{\Rsp(n_s)}{\Rsp'(n_s)n_s} \right|$ (dashed line).}
    \label{fig:my_label}
\end{figure}

For ``off-resonant Kerr laser," $\Rsp(n) \equiv 2g^2 \gamma_\perp/(\gamma_\perp^2 + (-\Delta_0 + \beta n)^2 )$. 
The gain of the cavity, defined as $G(n) = \Rsp(n) \Lambda/\left[\gamma_\parallel + \Rsp(n) n \right]$, depends non-monotonically on the light intensity (Fig. 3a). Therefore, the gain can intersect with the loss at one, two, or no point (with conditions for each scenario shown in Fig. 3b). In the case of two intersecting points, only the solution at the negative differential gain region is stable (Fig. 3a). The other solution is unstable under small perturbations because the positive differential gain will push the photon number further from its mean value. As such, we consider only the stable solution when analyzing the intensity noise. 

The noise behavior of the off-resonant Kerr laser is shown in Fig. 3c. When $\beta$ is negligible, the Fano-factor is $\kappa^2 \Delta_0^2/(2\Lambda g^2 \gamma_\perp)$, which is larger than the one of resonant linear laser with the same parameters even though the spontaneous emission rate in the off-resonant laser is suppressed. This effect can be explained by the overwhelming reduction of the mean photon number in the off-resonant linear laser due to the large detuning $\Delta_0$. As $\beta$ increases, there is a region where $\Rsp'$ is positive, unlike in the resonant case (Fig. 2c). Therefore, the noise first increases, reaching a value several orders of magnitude higher than the shot noise limit. In the sharply changing $\Rsp$ region, we can again estimate the noise intensity by Eq. \ref{eq:fano}. When the cavity and atoms are in resonance, or $\beta n_s - \Delta_0 = 0$, $\Rsp'$ is zero, thus we expect a local maximum value of $F$. When the total frequency shift is equal to the gain bandwidth ($\lvert \beta n_s - \Delta_0 \rvert = \gamma_\perp$), $F$ is minimized: at this point, the detuning value is small enough that it does not significantly reduce the steady-state light intensity. The minimum obtainable Factor factor in this laser architecture is $\gamma_\perp/(\gamma_\perp + \Delta_0)$. For the parameters we consider, this gives $F \lesssim 0.1$ (over 10 dB noise reduction). This minimum is realized when $\beta$ is set such that the cavity at equilibrium is detuned $\gamma_\perp$ lower than the atomic frequency. Such parameters can produce light which is more than 90$\%$ below the shot noise limt, but at photon number of $10^{12}$, which is clearly macroscopic.  

\section{Discussion and Outlook}

Here, we discuss some of the most important considerations relevant to the experimental observation of the effect we proposed. The most important element is that the per-photon nonlinear frequency shift $\beta$ be as large as possible compared to the polarization decay rate $\gamma_\perp.$ This ratio of parameters dictates the effective sharpness with which the gain drops off at higher photon numbers, leading to steady-state intensity noise reduction. Gain mediums Nd:YAG, Nd:YAP and Tm:YAG are all strong candidates due to their relatively narrow gain bandwidths set by $\gamma_\perp$. To maximize the nonlinear interaction $\beta$, one should use a material with low absorption and high $\chi^{(3)}$ such as GaP, and consider using a small cavity with high filling fraction of nonlinear material so that the intensity-dependent frequency shift is as large as possible. Note that while in principle, some of these materials (e.g., GaP) have two-photon absorption, it is quite weak, and may not be detrimental as such effects can lead to further number squeezing \cite{ritsch1990quantum, ritsch1992quantum,ispasoiu_two-photon_2002}. If two-photon absorption is not desired, then one can use a gain medium such as Er:YAG, Tm:YAG, or Tm:YAP, which have lasing wavelengths below the two-photon absorption cutoffs of Kerr materials such as GaAs and GaP. It is also worth pointing out that stronger Kerr shifts can be achieved at higher intensities. Thus, it is desirable to design a laser cavity that has highly reflective mirrors so that the intracavity power is orders of magnitude higher than the outcoupled power. The cavity parameters used in this work are consistent with those which can be realized experimentally. 

While for the sake of concreteness, we have based our discussions around a well-established solid-state laser geometry using a free space optical cavity, the concepts developed here should readily extend to other laser platforms, so long as the gain medium has a low bandwidth, and an appropriate amount of Kerr nonlinearity. Additional candidates for narrow bandwidth gain media could include gasses, engineered semiconductor, or quantum well transitions. On-chip platforms have the additional advantage that nanophotonic techniques can be used to engineer small mode volumes and efficient Kerr nonlinearity. 

In summary, we have proposed a laser architecture that uses sharply nonlinear gain to generate macroscopic sub-Poissonian light. We have discussed how this architecture can be implemented at optical frequencies using conventional solid state gain materials in conjunction with Kerr nonlinear materials. The parameters we have considered are experimentally accessible, and can lead to intra-cavity light which is 90\% below the shot noise limit. Importantly, the minimum achievable noise reduction is set purely by the parameters of the implementation (as opposed to mechanisms such as two-photon absorption, which have an intrinsic squeezing limit). Thus, with further optimization of the gain lineshape and Kerr nonlinearity on potentially different platforms, it may be possible to go far below the 10 dB reduction we discussed here, eventually leading to the generation of macroscopic near-Fock states. More broadly, this work points toward the further investigation of novel optical nonlinearities as a means to create quantum states of light. 

\section*{Acknowledgement}

This material is based upon
work supported in part by the Air Force Office of Scientific
Research under the award number FA9550-20-1-0115; the
work is also supported in part by the U. S. Army Research
Office through the Institute for Soldier Nanotechnologies at
MIT, under Collaborative Agreement Number W911NF-18-
2-0048. We also acknowledge support of Parviz Tayebati.

\bibliography{sample}

\providecommand{\noopsort}[1]{}\providecommand{\singleletter}[1]{#1}%
\begin{thebibliography}{42}%
\makeatletter
\providecommand \@ifxundefined [1]{%
 \@ifx{#1\undefined}
}%
\providecommand \@ifnum [1]{%
 \ifnum #1\expandafter \@firstoftwo
 \else \expandafter \@secondoftwo
 \fi
}%
\providecommand \@ifx [1]{%
 \ifx #1\expandafter \@firstoftwo
 \else \expandafter \@secondoftwo
 \fi
}%
\providecommand \natexlab [1]{#1}%
\providecommand \enquote  [1]{``#1''}%
\providecommand \bibnamefont  [1]{#1}%
\providecommand \bibfnamefont [1]{#1}%
\providecommand \citenamefont [1]{#1}%
\providecommand \href@noop [0]{\@secondoftwo}%
\providecommand \href [0]{\begingroup \@sanitize@url \@href}%
\providecommand \@href[1]{\@@startlink{#1}\@@href}%
\providecommand \@@href[1]{\endgroup#1\@@endlink}%
\providecommand \@sanitize@url [0]{\catcode `\\12\catcode `\$12\catcode
  `\&12\catcode `\#12\catcode `\^12\catcode `\_12\catcode `\%12\relax}%
\providecommand \@@startlink[1]{}%
\providecommand \@@endlink[0]{}%
\providecommand \url  [0]{\begingroup\@sanitize@url \@url }%
\providecommand \@url [1]{\endgroup\@href {#1}{\urlprefix }}%
\providecommand \urlprefix  [0]{URL }%
\providecommand \Eprint [0]{\href }%
\providecommand \doibase [0]{https://doi.org/}%
\providecommand \selectlanguage [0]{\@gobble}%
\providecommand \bibinfo  [0]{\@secondoftwo}%
\providecommand \bibfield  [0]{\@secondoftwo}%
\providecommand \translation [1]{[#1]}%
\providecommand \BibitemOpen [0]{}%
\providecommand \bibitemStop [0]{}%
\providecommand \bibitemNoStop [0]{.\EOS\space}%
\providecommand \EOS [0]{\spacefactor3000\relax}%
\providecommand \BibitemShut  [1]{\csname bibitem#1\endcsname}%
\let\auto@bib@innerbib\@empty
\bibitem [{\citenamefont {Davidovich}(1996)}]{davidovich1996sub}%
  \BibitemOpen
  \bibfield  {author} {\bibinfo {author} {\bibfnamefont {L.}~\bibnamefont
  {Davidovich}},\ }\bibfield  {title} {\bibinfo {title} {Sub-poissonian
  processes in quantum optics},\ }\href@noop {} {\bibfield  {journal} {\bibinfo
   {journal} {Reviews of Modern Physics}\ }\textbf {\bibinfo {volume} {68}},\
  \bibinfo {pages} {127} (\bibinfo {year} {1996})}\BibitemShut {NoStop}%
\bibitem [{\citenamefont {Thomas-Peter}\ \emph {et~al.}(2011)\citenamefont
  {Thomas-Peter}, \citenamefont {Smith}, \citenamefont {Datta}, \citenamefont
  {Zhang}, \citenamefont {Dorner},\ and\ \citenamefont
  {Walmsley}}]{thomas2011real}%
  \BibitemOpen
  \bibfield  {author} {\bibinfo {author} {\bibfnamefont {N.}~\bibnamefont
  {Thomas-Peter}}, \bibinfo {author} {\bibfnamefont {B.~J.}\ \bibnamefont
  {Smith}}, \bibinfo {author} {\bibfnamefont {A.}~\bibnamefont {Datta}},
  \bibinfo {author} {\bibfnamefont {L.}~\bibnamefont {Zhang}}, \bibinfo
  {author} {\bibfnamefont {U.}~\bibnamefont {Dorner}},\ and\ \bibinfo {author}
  {\bibfnamefont {I.~A.}\ \bibnamefont {Walmsley}},\ }\bibfield  {title}
  {\bibinfo {title} {Real-world quantum sensors: evaluating resources for
  precision measurement},\ }\href@noop {} {\bibfield  {journal} {\bibinfo
  {journal} {Physical review letters}\ }\textbf {\bibinfo {volume} {107}},\
  \bibinfo {pages} {113603} (\bibinfo {year} {2011})}\BibitemShut {NoStop}%
\bibitem [{\citenamefont {Teich}\ and\ \citenamefont
  {Saleh}(1989)}]{teich1989squeezed}%
  \BibitemOpen
  \bibfield  {author} {\bibinfo {author} {\bibfnamefont {M.~C.}\ \bibnamefont
  {Teich}}\ and\ \bibinfo {author} {\bibfnamefont {B.~E.}\ \bibnamefont
  {Saleh}},\ }\bibfield  {title} {\bibinfo {title} {Squeezed state of light},\
  }\href@noop {} {\bibfield  {journal} {\bibinfo  {journal} {Quantum Optics:
  Journal of the European Optical Society Part B}\ }\textbf {\bibinfo {volume}
  {1}},\ \bibinfo {pages} {153} (\bibinfo {year} {1989})}\BibitemShut {NoStop}%
\bibitem [{\citenamefont {Aaronson}\ and\ \citenamefont
  {Arkhipov}(2011)}]{aaronson2011computational}%
  \BibitemOpen
  \bibfield  {author} {\bibinfo {author} {\bibfnamefont {S.}~\bibnamefont
  {Aaronson}}\ and\ \bibinfo {author} {\bibfnamefont {A.}~\bibnamefont
  {Arkhipov}},\ }\bibfield  {title} {\bibinfo {title} {The computational
  complexity of linear optics},\ }in\ \href@noop {} {\emph {\bibinfo
  {booktitle} {Proceedings of the forty-third annual ACM symposium on Theory of
  computing}}}\ (\bibinfo {year} {2011})\ pp.\ \bibinfo {pages}
  {333--342}\BibitemShut {NoStop}%
\bibitem [{\citenamefont {Lund}\ \emph {et~al.}(2014)\citenamefont {Lund},
  \citenamefont {Laing}, \citenamefont {Rahimi-Keshari}, \citenamefont
  {Rudolph}, \citenamefont {O’Brien},\ and\ \citenamefont
  {Ralph}}]{lund2014boson}%
  \BibitemOpen
  \bibfield  {author} {\bibinfo {author} {\bibfnamefont {A.~P.}\ \bibnamefont
  {Lund}}, \bibinfo {author} {\bibfnamefont {A.}~\bibnamefont {Laing}},
  \bibinfo {author} {\bibfnamefont {S.}~\bibnamefont {Rahimi-Keshari}},
  \bibinfo {author} {\bibfnamefont {T.}~\bibnamefont {Rudolph}}, \bibinfo
  {author} {\bibfnamefont {J.~L.}\ \bibnamefont {O’Brien}},\ and\ \bibinfo
  {author} {\bibfnamefont {T.~C.}\ \bibnamefont {Ralph}},\ }\bibfield  {title}
  {\bibinfo {title} {Boson sampling from a gaussian state},\ }\href@noop {}
  {\bibfield  {journal} {\bibinfo  {journal} {Physical review letters}\
  }\textbf {\bibinfo {volume} {113}},\ \bibinfo {pages} {100502} (\bibinfo
  {year} {2014})}\BibitemShut {NoStop}%
\bibitem [{\citenamefont {Huh}\ \emph {et~al.}(2015)\citenamefont {Huh},
  \citenamefont {Guerreschi}, \citenamefont {Peropadre}, \citenamefont
  {McClean},\ and\ \citenamefont {Aspuru-Guzik}}]{huh2015boson}%
  \BibitemOpen
  \bibfield  {author} {\bibinfo {author} {\bibfnamefont {J.}~\bibnamefont
  {Huh}}, \bibinfo {author} {\bibfnamefont {G.~G.}\ \bibnamefont {Guerreschi}},
  \bibinfo {author} {\bibfnamefont {B.}~\bibnamefont {Peropadre}}, \bibinfo
  {author} {\bibfnamefont {J.~R.}\ \bibnamefont {McClean}},\ and\ \bibinfo
  {author} {\bibfnamefont {A.}~\bibnamefont {Aspuru-Guzik}},\ }\bibfield
  {title} {\bibinfo {title} {Boson sampling for molecular vibronic spectra},\
  }\href@noop {} {\bibfield  {journal} {\bibinfo  {journal} {Nature Photonics}\
  }\textbf {\bibinfo {volume} {9}},\ \bibinfo {pages} {615} (\bibinfo {year}
  {2015})}\BibitemShut {NoStop}%
\bibitem [{\citenamefont {Wang}\ \emph {et~al.}(2017)\citenamefont {Wang},
  \citenamefont {He}, \citenamefont {Li}, \citenamefont {Su}, \citenamefont
  {Li}, \citenamefont {Huang}, \citenamefont {Ding}, \citenamefont {Chen},
  \citenamefont {Liu}, \citenamefont {Qin} \emph {et~al.}}]{wang2017high}%
  \BibitemOpen
  \bibfield  {author} {\bibinfo {author} {\bibfnamefont {H.}~\bibnamefont
  {Wang}}, \bibinfo {author} {\bibfnamefont {Y.}~\bibnamefont {He}}, \bibinfo
  {author} {\bibfnamefont {Y.-H.}\ \bibnamefont {Li}}, \bibinfo {author}
  {\bibfnamefont {Z.-E.}\ \bibnamefont {Su}}, \bibinfo {author} {\bibfnamefont
  {B.}~\bibnamefont {Li}}, \bibinfo {author} {\bibfnamefont {H.-L.}\
  \bibnamefont {Huang}}, \bibinfo {author} {\bibfnamefont {X.}~\bibnamefont
  {Ding}}, \bibinfo {author} {\bibfnamefont {M.-C.}\ \bibnamefont {Chen}},
  \bibinfo {author} {\bibfnamefont {C.}~\bibnamefont {Liu}}, \bibinfo {author}
  {\bibfnamefont {J.}~\bibnamefont {Qin}}, \emph {et~al.},\ }\bibfield  {title}
  {\bibinfo {title} {High-efficiency multiphoton boson sampling},\ }\href@noop
  {} {\bibfield  {journal} {\bibinfo  {journal} {Nature Photonics}\ }\textbf
  {\bibinfo {volume} {11}},\ \bibinfo {pages} {361} (\bibinfo {year}
  {2017})}\BibitemShut {NoStop}%
\bibitem [{\citenamefont {Hamilton}\ \emph {et~al.}(2017)\citenamefont
  {Hamilton}, \citenamefont {Kruse}, \citenamefont {Sansoni}, \citenamefont
  {Barkhofen}, \citenamefont {Silberhorn},\ and\ \citenamefont
  {Jex}}]{hamilton2017gaussian}%
  \BibitemOpen
  \bibfield  {author} {\bibinfo {author} {\bibfnamefont {C.~S.}\ \bibnamefont
  {Hamilton}}, \bibinfo {author} {\bibfnamefont {R.}~\bibnamefont {Kruse}},
  \bibinfo {author} {\bibfnamefont {L.}~\bibnamefont {Sansoni}}, \bibinfo
  {author} {\bibfnamefont {S.}~\bibnamefont {Barkhofen}}, \bibinfo {author}
  {\bibfnamefont {C.}~\bibnamefont {Silberhorn}},\ and\ \bibinfo {author}
  {\bibfnamefont {I.}~\bibnamefont {Jex}},\ }\bibfield  {title} {\bibinfo
  {title} {Gaussian boson sampling},\ }\href@noop {} {\bibfield  {journal}
  {\bibinfo  {journal} {Physical review letters}\ }\textbf {\bibinfo {volume}
  {119}},\ \bibinfo {pages} {170501} (\bibinfo {year} {2017})}\BibitemShut
  {NoStop}%
\bibitem [{\citenamefont {Brod}\ \emph {et~al.}(2019)\citenamefont {Brod},
  \citenamefont {Galv{\~a}o}, \citenamefont {Crespi}, \citenamefont {Osellame},
  \citenamefont {Spagnolo},\ and\ \citenamefont
  {Sciarrino}}]{brod2019photonic}%
  \BibitemOpen
  \bibfield  {author} {\bibinfo {author} {\bibfnamefont {D.~J.}\ \bibnamefont
  {Brod}}, \bibinfo {author} {\bibfnamefont {E.~F.}\ \bibnamefont
  {Galv{\~a}o}}, \bibinfo {author} {\bibfnamefont {A.}~\bibnamefont {Crespi}},
  \bibinfo {author} {\bibfnamefont {R.}~\bibnamefont {Osellame}}, \bibinfo
  {author} {\bibfnamefont {N.}~\bibnamefont {Spagnolo}},\ and\ \bibinfo
  {author} {\bibfnamefont {F.}~\bibnamefont {Sciarrino}},\ }\bibfield  {title}
  {\bibinfo {title} {Photonic implementation of boson sampling: a review},\
  }\href@noop {} {\bibfield  {journal} {\bibinfo  {journal} {Advanced
  Photonics}\ }\textbf {\bibinfo {volume} {1}},\ \bibinfo {pages} {034001}
  (\bibinfo {year} {2019})}\BibitemShut {NoStop}%
\bibitem [{\citenamefont {Wang}\ \emph {et~al.}(2020)\citenamefont {Wang},
  \citenamefont {Curtis}, \citenamefont {Lester}, \citenamefont {Zhang},
  \citenamefont {Gao}, \citenamefont {Freeze}, \citenamefont {Batista},
  \citenamefont {Vaccaro}, \citenamefont {Chuang}, \citenamefont {Frunzio}
  \emph {et~al.}}]{wang2020efficient}%
  \BibitemOpen
  \bibfield  {author} {\bibinfo {author} {\bibfnamefont {C.~S.}\ \bibnamefont
  {Wang}}, \bibinfo {author} {\bibfnamefont {J.~C.}\ \bibnamefont {Curtis}},
  \bibinfo {author} {\bibfnamefont {B.~J.}\ \bibnamefont {Lester}}, \bibinfo
  {author} {\bibfnamefont {Y.}~\bibnamefont {Zhang}}, \bibinfo {author}
  {\bibfnamefont {Y.~Y.}\ \bibnamefont {Gao}}, \bibinfo {author} {\bibfnamefont
  {J.}~\bibnamefont {Freeze}}, \bibinfo {author} {\bibfnamefont {V.~S.}\
  \bibnamefont {Batista}}, \bibinfo {author} {\bibfnamefont {P.~H.}\
  \bibnamefont {Vaccaro}}, \bibinfo {author} {\bibfnamefont {I.~L.}\
  \bibnamefont {Chuang}}, \bibinfo {author} {\bibfnamefont {L.}~\bibnamefont
  {Frunzio}}, \emph {et~al.},\ }\bibfield  {title} {\bibinfo {title} {Efficient
  multiphoton sampling of molecular vibronic spectra on a superconducting
  bosonic processor},\ }\href@noop {} {\bibfield  {journal} {\bibinfo
  {journal} {Physical Review X}\ }\textbf {\bibinfo {volume} {10}},\ \bibinfo
  {pages} {021060} (\bibinfo {year} {2020})}\BibitemShut {NoStop}%
\bibitem [{\citenamefont {Rempe}\ \emph {et~al.}(1990)\citenamefont {Rempe},
  \citenamefont {Schmidt-Kaler},\ and\ \citenamefont
  {Walther}}]{rempe1990observation}%
  \BibitemOpen
  \bibfield  {author} {\bibinfo {author} {\bibfnamefont {G.}~\bibnamefont
  {Rempe}}, \bibinfo {author} {\bibfnamefont {F.}~\bibnamefont
  {Schmidt-Kaler}},\ and\ \bibinfo {author} {\bibfnamefont {H.}~\bibnamefont
  {Walther}},\ }\bibfield  {title} {\bibinfo {title} {Observation of
  sub-poissonian photon statistics in a micromaser},\ }\href@noop {} {\bibfield
   {journal} {\bibinfo  {journal} {Physical review letters}\ }\textbf {\bibinfo
  {volume} {64}},\ \bibinfo {pages} {2783} (\bibinfo {year}
  {1990})}\BibitemShut {NoStop}%
\bibitem [{\citenamefont {Varcoe}\ \emph {et~al.}(2000)\citenamefont {Varcoe},
  \citenamefont {Brattke}, \citenamefont {Weidinger},\ and\ \citenamefont
  {Walther}}]{varcoe2000preparing}%
  \BibitemOpen
  \bibfield  {author} {\bibinfo {author} {\bibfnamefont {B.~T.}\ \bibnamefont
  {Varcoe}}, \bibinfo {author} {\bibfnamefont {S.}~\bibnamefont {Brattke}},
  \bibinfo {author} {\bibfnamefont {M.}~\bibnamefont {Weidinger}},\ and\
  \bibinfo {author} {\bibfnamefont {H.}~\bibnamefont {Walther}},\ }\bibfield
  {title} {\bibinfo {title} {Preparing pure photon number states of the
  radiation field},\ }\href@noop {} {\bibfield  {journal} {\bibinfo  {journal}
  {Nature}\ }\textbf {\bibinfo {volume} {403}},\ \bibinfo {pages} {743}
  (\bibinfo {year} {2000})}\BibitemShut {NoStop}%
\bibitem [{\citenamefont {Sayrin}\ \emph {et~al.}(2011)\citenamefont {Sayrin},
  \citenamefont {Dotsenko}, \citenamefont {Zhou}, \citenamefont {Peaudecerf},
  \citenamefont {Rybarczyk}, \citenamefont {Gleyzes}, \citenamefont {Rouchon},
  \citenamefont {Mirrahimi}, \citenamefont {Amini}, \citenamefont {Brune} \emph
  {et~al.}}]{sayrin2011real}%
  \BibitemOpen
  \bibfield  {author} {\bibinfo {author} {\bibfnamefont {C.}~\bibnamefont
  {Sayrin}}, \bibinfo {author} {\bibfnamefont {I.}~\bibnamefont {Dotsenko}},
  \bibinfo {author} {\bibfnamefont {X.}~\bibnamefont {Zhou}}, \bibinfo {author}
  {\bibfnamefont {B.}~\bibnamefont {Peaudecerf}}, \bibinfo {author}
  {\bibfnamefont {T.}~\bibnamefont {Rybarczyk}}, \bibinfo {author}
  {\bibfnamefont {S.}~\bibnamefont {Gleyzes}}, \bibinfo {author} {\bibfnamefont
  {P.}~\bibnamefont {Rouchon}}, \bibinfo {author} {\bibfnamefont
  {M.}~\bibnamefont {Mirrahimi}}, \bibinfo {author} {\bibfnamefont
  {H.}~\bibnamefont {Amini}}, \bibinfo {author} {\bibfnamefont
  {M.}~\bibnamefont {Brune}}, \emph {et~al.},\ }\bibfield  {title} {\bibinfo
  {title} {Real-time quantum feedback prepares and stabilizes photon number
  states},\ }\href@noop {} {\bibfield  {journal} {\bibinfo  {journal} {Nature}\
  }\textbf {\bibinfo {volume} {477}},\ \bibinfo {pages} {73} (\bibinfo {year}
  {2011})}\BibitemShut {NoStop}%
\bibitem [{\citenamefont {Uria}\ \emph {et~al.}(2020)\citenamefont {Uria},
  \citenamefont {Solano},\ and\ \citenamefont
  {Hermann-Avigliano}}]{uria2020deterministic}%
  \BibitemOpen
  \bibfield  {author} {\bibinfo {author} {\bibfnamefont {M.}~\bibnamefont
  {Uria}}, \bibinfo {author} {\bibfnamefont {P.}~\bibnamefont {Solano}},\ and\
  \bibinfo {author} {\bibfnamefont {C.}~\bibnamefont {Hermann-Avigliano}},\
  }\bibfield  {title} {\bibinfo {title} {Deterministic generation of large fock
  states},\ }\href@noop {} {\bibfield  {journal} {\bibinfo  {journal} {Physical
  Review Letters}\ }\textbf {\bibinfo {volume} {125}},\ \bibinfo {pages}
  {093603} (\bibinfo {year} {2020})}\BibitemShut {NoStop}%
\bibitem [{\citenamefont {Canela}\ and\ \citenamefont
  {Carmichael}(2020)}]{canela2020bright}%
  \BibitemOpen
  \bibfield  {author} {\bibinfo {author} {\bibfnamefont {V.}~\bibnamefont
  {Canela}}\ and\ \bibinfo {author} {\bibfnamefont {H.}~\bibnamefont
  {Carmichael}},\ }\bibfield  {title} {\bibinfo {title} {Bright sub-poissonian
  light through intrinsic feedback and external control},\ }\href@noop {}
  {\bibfield  {journal} {\bibinfo  {journal} {Physical review letters}\
  }\textbf {\bibinfo {volume} {124}},\ \bibinfo {pages} {063604} (\bibinfo
  {year} {2020})}\BibitemShut {NoStop}%
\bibitem [{\citenamefont {Hofheinz}\ \emph {et~al.}(2008)\citenamefont
  {Hofheinz}, \citenamefont {Weig}, \citenamefont {Ansmann}, \citenamefont
  {Bialczak}, \citenamefont {Lucero}, \citenamefont {Neeley}, \citenamefont
  {O’connell}, \citenamefont {Wang}, \citenamefont {Martinis},\ and\
  \citenamefont {Cleland}}]{hofheinz2008generation}%
  \BibitemOpen
  \bibfield  {author} {\bibinfo {author} {\bibfnamefont {M.}~\bibnamefont
  {Hofheinz}}, \bibinfo {author} {\bibfnamefont {E.}~\bibnamefont {Weig}},
  \bibinfo {author} {\bibfnamefont {M.}~\bibnamefont {Ansmann}}, \bibinfo
  {author} {\bibfnamefont {R.~C.}\ \bibnamefont {Bialczak}}, \bibinfo {author}
  {\bibfnamefont {E.}~\bibnamefont {Lucero}}, \bibinfo {author} {\bibfnamefont
  {M.}~\bibnamefont {Neeley}}, \bibinfo {author} {\bibfnamefont
  {A.}~\bibnamefont {O’connell}}, \bibinfo {author} {\bibfnamefont
  {H.}~\bibnamefont {Wang}}, \bibinfo {author} {\bibfnamefont {J.~M.}\
  \bibnamefont {Martinis}},\ and\ \bibinfo {author} {\bibfnamefont
  {A.}~\bibnamefont {Cleland}},\ }\bibfield  {title} {\bibinfo {title}
  {Generation of fock states in a superconducting quantum circuit},\
  }\href@noop {} {\bibfield  {journal} {\bibinfo  {journal} {Nature}\ }\textbf
  {\bibinfo {volume} {454}},\ \bibinfo {pages} {310} (\bibinfo {year}
  {2008})}\BibitemShut {NoStop}%
\bibitem [{\citenamefont {Heeres}\ \emph {et~al.}(2017)\citenamefont {Heeres},
  \citenamefont {Reinhold}, \citenamefont {Ofek}, \citenamefont {Frunzio},
  \citenamefont {Jiang}, \citenamefont {Devoret},\ and\ \citenamefont
  {Schoelkopf}}]{heeres2017implementing}%
  \BibitemOpen
  \bibfield  {author} {\bibinfo {author} {\bibfnamefont {R.~W.}\ \bibnamefont
  {Heeres}}, \bibinfo {author} {\bibfnamefont {P.}~\bibnamefont {Reinhold}},
  \bibinfo {author} {\bibfnamefont {N.}~\bibnamefont {Ofek}}, \bibinfo {author}
  {\bibfnamefont {L.}~\bibnamefont {Frunzio}}, \bibinfo {author} {\bibfnamefont
  {L.}~\bibnamefont {Jiang}}, \bibinfo {author} {\bibfnamefont {M.~H.}\
  \bibnamefont {Devoret}},\ and\ \bibinfo {author} {\bibfnamefont {R.~J.}\
  \bibnamefont {Schoelkopf}},\ }\bibfield  {title} {\bibinfo {title}
  {Implementing a universal gate set on a logical qubit encoded in an
  oscillator},\ }\href@noop {} {\bibfield  {journal} {\bibinfo  {journal}
  {Nature communications}\ }\textbf {\bibinfo {volume} {8}},\ \bibinfo {pages}
  {1} (\bibinfo {year} {2017})}\BibitemShut {NoStop}%
\bibitem [{\citenamefont {Birnbaum}\ \emph {et~al.}(2005)\citenamefont
  {Birnbaum}, \citenamefont {Boca}, \citenamefont {Miller}, \citenamefont
  {Boozer}, \citenamefont {Northup},\ and\ \citenamefont
  {Kimble}}]{birnbaum2005photon}%
  \BibitemOpen
  \bibfield  {author} {\bibinfo {author} {\bibfnamefont {K.~M.}\ \bibnamefont
  {Birnbaum}}, \bibinfo {author} {\bibfnamefont {A.}~\bibnamefont {Boca}},
  \bibinfo {author} {\bibfnamefont {R.}~\bibnamefont {Miller}}, \bibinfo
  {author} {\bibfnamefont {A.~D.}\ \bibnamefont {Boozer}}, \bibinfo {author}
  {\bibfnamefont {T.~E.}\ \bibnamefont {Northup}},\ and\ \bibinfo {author}
  {\bibfnamefont {H.~J.}\ \bibnamefont {Kimble}},\ }\bibfield  {title}
  {\bibinfo {title} {Photon blockade in an optical cavity with one trapped
  atom},\ }\href@noop {} {\bibfield  {journal} {\bibinfo  {journal} {Nature}\
  }\textbf {\bibinfo {volume} {436}},\ \bibinfo {pages} {87} (\bibinfo {year}
  {2005})}\BibitemShut {NoStop}%
\bibitem [{\citenamefont {Flayac}\ and\ \citenamefont
  {Savona}(2017)}]{flayac2017unconventional}%
  \BibitemOpen
  \bibfield  {author} {\bibinfo {author} {\bibfnamefont {H.}~\bibnamefont
  {Flayac}}\ and\ \bibinfo {author} {\bibfnamefont {V.}~\bibnamefont
  {Savona}},\ }\bibfield  {title} {\bibinfo {title} {Unconventional photon
  blockade},\ }\href@noop {} {\bibfield  {journal} {\bibinfo  {journal}
  {Physical Review A}\ }\textbf {\bibinfo {volume} {96}},\ \bibinfo {pages}
  {053810} (\bibinfo {year} {2017})}\BibitemShut {NoStop}%
\bibitem [{\citenamefont {Rivera}\ \emph {et~al.}(2021)\citenamefont {Rivera},
  \citenamefont {Sloan}, \citenamefont {Kaminer},\ and\ \citenamefont
  {Soljacic}}]{Rivera2021Fock}%
  \BibitemOpen
  \bibfield  {author} {\bibinfo {author} {\bibfnamefont {N.}~\bibnamefont
  {Rivera}}, \bibinfo {author} {\bibfnamefont {J.}~\bibnamefont {Sloan}},
  \bibinfo {author} {\bibfnamefont {I.}~\bibnamefont {Kaminer}},\ and\ \bibinfo
  {author} {\bibfnamefont {M.}~\bibnamefont {Soljacic}},\ }\href@noop {}
  {\bibinfo {title} {Fock lasers based on deep-strong coupling of light and
  matter}} (\bibinfo {year} {2021}),\ \Eprint
  {https://arxiv.org/abs/2111.07010} {arXiv:2111.07010 [quant-ph]} \BibitemShut
  {NoStop}%
\bibitem [{\citenamefont {Walls}\ \emph {et~al.}(1990)\citenamefont {Walls},
  \citenamefont {Collett},\ and\ \citenamefont {Lane}}]{two_photon_absoprtion}%
  \BibitemOpen
  \bibfield  {author} {\bibinfo {author} {\bibfnamefont {D.~F.}\ \bibnamefont
  {Walls}}, \bibinfo {author} {\bibfnamefont {M.~J.}\ \bibnamefont {Collett}},\
  and\ \bibinfo {author} {\bibfnamefont {A.~S.}\ \bibnamefont {Lane}},\
  }\bibfield  {title} {\bibinfo {title} {Amplitude-noise reduction in lasers
  with intracavity nonlinear elements},\ }\href@noop {} {\bibfield  {journal}
  {\bibinfo  {journal} {Phys. Rev. A}\ }\textbf {\bibinfo {volume} {42}},\
  \bibinfo {pages} {4366} (\bibinfo {year} {1990})}\BibitemShut {NoStop}%
\bibitem [{\citenamefont {Ritsch}(1990)}]{ritsch1990quantum}%
  \BibitemOpen
  \bibfield  {author} {\bibinfo {author} {\bibfnamefont {H.}~\bibnamefont
  {Ritsch}},\ }\bibfield  {title} {\bibinfo {title} {Quantum noise reduction in
  lasers with nonlinear absorbers},\ }\href@noop {} {\bibfield  {journal}
  {\bibinfo  {journal} {Quantum Optics: Journal of the European Optical Society
  Part B}\ }\textbf {\bibinfo {volume} {2}},\ \bibinfo {pages} {189} (\bibinfo
  {year} {1990})}\BibitemShut {NoStop}%
\bibitem [{\citenamefont {Ritsch}\ \emph {et~al.}(1992)\citenamefont {Ritsch},
  \citenamefont {Marte},\ and\ \citenamefont {Zoller}}]{ritsch1992quantum}%
  \BibitemOpen
  \bibfield  {author} {\bibinfo {author} {\bibfnamefont {H.}~\bibnamefont
  {Ritsch}}, \bibinfo {author} {\bibfnamefont {M.}~\bibnamefont {Marte}},\ and\
  \bibinfo {author} {\bibfnamefont {P.}~\bibnamefont {Zoller}},\ }\bibfield
  {title} {\bibinfo {title} {Quantum noise reduction in raman lasers},\
  }\href@noop {} {\bibfield  {journal} {\bibinfo  {journal} {EPL (Europhysics
  Letters)}\ }\textbf {\bibinfo {volume} {19}},\ \bibinfo {pages} {7} (\bibinfo
  {year} {1992})}\BibitemShut {NoStop}%
\bibitem [{\citenamefont {Kitagawa}\ and\ \citenamefont
  {Yamamoto}(1986)}]{kitagawa1986number}%
  \BibitemOpen
  \bibfield  {author} {\bibinfo {author} {\bibfnamefont {M.}~\bibnamefont
  {Kitagawa}}\ and\ \bibinfo {author} {\bibfnamefont {Y.}~\bibnamefont
  {Yamamoto}},\ }\bibfield  {title} {\bibinfo {title} {Number-phase
  minimum-uncertainty state with reduced number uncertainty in a kerr nonlinear
  interferometer},\ }\href@noop {} {\bibfield  {journal} {\bibinfo  {journal}
  {Physical Review A}\ }\textbf {\bibinfo {volume} {34}},\ \bibinfo {pages}
  {3974} (\bibinfo {year} {1986})}\BibitemShut {NoStop}%
\bibitem [{\citenamefont {Shirasaki}\ and\ \citenamefont
  {Haus}(1990)}]{shirasaki1990squeezing}%
  \BibitemOpen
  \bibfield  {author} {\bibinfo {author} {\bibfnamefont {M.}~\bibnamefont
  {Shirasaki}}\ and\ \bibinfo {author} {\bibfnamefont {H.~A.}\ \bibnamefont
  {Haus}},\ }\bibfield  {title} {\bibinfo {title} {Squeezing of pulses in a
  nonlinear interferometer},\ }\href@noop {} {\bibfield  {journal} {\bibinfo
  {journal} {JOSA B}\ }\textbf {\bibinfo {volume} {7}},\ \bibinfo {pages} {30}
  (\bibinfo {year} {1990})}\BibitemShut {NoStop}%
\bibitem [{\citenamefont {Bergman}\ and\ \citenamefont
  {Haus}(1991)}]{bergman1991squeezing}%
  \BibitemOpen
  \bibfield  {author} {\bibinfo {author} {\bibfnamefont {K.}~\bibnamefont
  {Bergman}}\ and\ \bibinfo {author} {\bibfnamefont {H.}~\bibnamefont {Haus}},\
  }\bibfield  {title} {\bibinfo {title} {Squeezing in fibers with optical
  pulses},\ }\href@noop {} {\bibfield  {journal} {\bibinfo  {journal} {Optics
  letters}\ }\textbf {\bibinfo {volume} {16}},\ \bibinfo {pages} {663}
  (\bibinfo {year} {1991})}\BibitemShut {NoStop}%
\bibitem [{\citenamefont {Kitching}\ \emph {et~al.}(1994)\citenamefont
  {Kitching}, \citenamefont {Boyd}, \citenamefont {Yariv},\ and\ \citenamefont
  {Shevy}}]{kitching1994amplitude}%
  \BibitemOpen
  \bibfield  {author} {\bibinfo {author} {\bibfnamefont {J.}~\bibnamefont
  {Kitching}}, \bibinfo {author} {\bibfnamefont {R.}~\bibnamefont {Boyd}},
  \bibinfo {author} {\bibfnamefont {A.}~\bibnamefont {Yariv}},\ and\ \bibinfo
  {author} {\bibfnamefont {Y.}~\bibnamefont {Shevy}},\ }\bibfield  {title}
  {\bibinfo {title} {Amplitude noise reduction in semiconductor lasers with
  weak, dispersive optical feedback},\ }\href@noop {} {\bibfield  {journal}
  {\bibinfo  {journal} {Optics letters}\ }\textbf {\bibinfo {volume} {19}},\
  \bibinfo {pages} {1331} (\bibinfo {year} {1994})}\BibitemShut {NoStop}%
\bibitem [{\citenamefont {Schmitt}\ \emph {et~al.}(1998)\citenamefont
  {Schmitt}, \citenamefont {Ficker}, \citenamefont {Wolff}, \citenamefont
  {K{\"o}nig}, \citenamefont {Sizmann},\ and\ \citenamefont
  {Leuchs}}]{schmitt1998photon}%
  \BibitemOpen
  \bibfield  {author} {\bibinfo {author} {\bibfnamefont {S.}~\bibnamefont
  {Schmitt}}, \bibinfo {author} {\bibfnamefont {J.}~\bibnamefont {Ficker}},
  \bibinfo {author} {\bibfnamefont {M.}~\bibnamefont {Wolff}}, \bibinfo
  {author} {\bibfnamefont {F.}~\bibnamefont {K{\"o}nig}}, \bibinfo {author}
  {\bibfnamefont {A.}~\bibnamefont {Sizmann}},\ and\ \bibinfo {author}
  {\bibfnamefont {G.}~\bibnamefont {Leuchs}},\ }\bibfield  {title} {\bibinfo
  {title} {Photon-number squeezed solitons from an asymmetric fiber-optic
  sagnac interferometer},\ }\href@noop {} {\bibfield  {journal} {\bibinfo
  {journal} {Physical review letters}\ }\textbf {\bibinfo {volume} {81}},\
  \bibinfo {pages} {2446} (\bibinfo {year} {1998})}\BibitemShut {NoStop}%
\bibitem [{\citenamefont {Andersen}\ \emph {et~al.}(2016)\citenamefont
  {Andersen}, \citenamefont {Gehring}, \citenamefont {Marquardt},\ and\
  \citenamefont {Leuchs}}]{andersen201630}%
  \BibitemOpen
  \bibfield  {author} {\bibinfo {author} {\bibfnamefont {U.~L.}\ \bibnamefont
  {Andersen}}, \bibinfo {author} {\bibfnamefont {T.}~\bibnamefont {Gehring}},
  \bibinfo {author} {\bibfnamefont {C.}~\bibnamefont {Marquardt}},\ and\
  \bibinfo {author} {\bibfnamefont {G.}~\bibnamefont {Leuchs}},\ }\bibfield
  {title} {\bibinfo {title} {30 years of squeezed light generation},\
  }\href@noop {} {\bibfield  {journal} {\bibinfo  {journal} {Physica Scripta}\
  }\textbf {\bibinfo {volume} {91}},\ \bibinfo {pages} {053001} (\bibinfo
  {year} {2016})}\BibitemShut {NoStop}%
\bibitem [{\citenamefont {Gonz{\'a}lez-Tudela}\ \emph
  {et~al.}(2017)\citenamefont {Gonz{\'a}lez-Tudela}, \citenamefont {Paulisch},
  \citenamefont {Kimble},\ and\ \citenamefont {Cirac}}]{gonzalez2017efficient}%
  \BibitemOpen
  \bibfield  {author} {\bibinfo {author} {\bibfnamefont {A.}~\bibnamefont
  {Gonz{\'a}lez-Tudela}}, \bibinfo {author} {\bibfnamefont {V.}~\bibnamefont
  {Paulisch}}, \bibinfo {author} {\bibfnamefont {H.}~\bibnamefont {Kimble}},\
  and\ \bibinfo {author} {\bibfnamefont {J.~I.}\ \bibnamefont {Cirac}},\
  }\bibfield  {title} {\bibinfo {title} {Efficient multiphoton generation in
  waveguide quantum electrodynamics},\ }\href@noop {} {\bibfield  {journal}
  {\bibinfo  {journal} {Physical Review Letters}\ }\textbf {\bibinfo {volume}
  {118}},\ \bibinfo {pages} {213601} (\bibinfo {year} {2017})}\BibitemShut
  {NoStop}%
\bibitem [{\citenamefont {Kilin}\ and\ \citenamefont
  {Horoshko}(1995)}]{kilin1995fock}%
  \BibitemOpen
  \bibfield  {author} {\bibinfo {author} {\bibfnamefont {S.~Y.}\ \bibnamefont
  {Kilin}}\ and\ \bibinfo {author} {\bibfnamefont {D.}~\bibnamefont
  {Horoshko}},\ }\bibfield  {title} {\bibinfo {title} {Fock state generation by
  the methods of nonlinear optics},\ }\href@noop {} {\bibfield  {journal}
  {\bibinfo  {journal} {Physical review letters}\ }\textbf {\bibinfo {volume}
  {74}},\ \bibinfo {pages} {5206} (\bibinfo {year} {1995})}\BibitemShut
  {NoStop}%
\bibitem [{\citenamefont {Leo{\'n}ski}\ \emph {et~al.}(1997)\citenamefont
  {Leo{\'n}ski}, \citenamefont {Dyrting},\ and\ \citenamefont
  {Tana{\'s}}}]{leonski1997fock}%
  \BibitemOpen
  \bibfield  {author} {\bibinfo {author} {\bibfnamefont {W.}~\bibnamefont
  {Leo{\'n}ski}}, \bibinfo {author} {\bibfnamefont {S.}~\bibnamefont
  {Dyrting}},\ and\ \bibinfo {author} {\bibfnamefont {R.}~\bibnamefont
  {Tana{\'s}}},\ }\bibfield  {title} {\bibinfo {title} {Fock states generation
  in a kicked cavity with a nonlinear medium},\ }\href@noop {} {\bibfield
  {journal} {\bibinfo  {journal} {journal of modern optics}\ }\textbf {\bibinfo
  {volume} {44}},\ \bibinfo {pages} {2105} (\bibinfo {year}
  {1997})}\BibitemShut {NoStop}%
\bibitem [{\citenamefont {Mu{\~n}oz}\ \emph {et~al.}(2014)\citenamefont
  {Mu{\~n}oz}, \citenamefont {Del~Valle}, \citenamefont {Tudela}, \citenamefont
  {M{\"u}ller}, \citenamefont {Lichtmannecker}, \citenamefont {Kaniber},
  \citenamefont {Tejedor}, \citenamefont {Finley},\ and\ \citenamefont
  {Laussy}}]{munoz2014emitters}%
  \BibitemOpen
  \bibfield  {author} {\bibinfo {author} {\bibfnamefont {C.~S.}\ \bibnamefont
  {Mu{\~n}oz}}, \bibinfo {author} {\bibfnamefont {E.}~\bibnamefont
  {Del~Valle}}, \bibinfo {author} {\bibfnamefont {A.~G.}\ \bibnamefont
  {Tudela}}, \bibinfo {author} {\bibfnamefont {K.}~\bibnamefont {M{\"u}ller}},
  \bibinfo {author} {\bibfnamefont {S.}~\bibnamefont {Lichtmannecker}},
  \bibinfo {author} {\bibfnamefont {M.}~\bibnamefont {Kaniber}}, \bibinfo
  {author} {\bibfnamefont {C.}~\bibnamefont {Tejedor}}, \bibinfo {author}
  {\bibfnamefont {J.}~\bibnamefont {Finley}},\ and\ \bibinfo {author}
  {\bibfnamefont {F.}~\bibnamefont {Laussy}},\ }\bibfield  {title} {\bibinfo
  {title} {Emitters of n-photon bundles},\ }\href@noop {} {\bibfield  {journal}
  {\bibinfo  {journal} {Nature photonics}\ }\textbf {\bibinfo {volume} {8}},\
  \bibinfo {pages} {550} (\bibinfo {year} {2014})}\BibitemShut {NoStop}%
\bibitem [{\citenamefont {Yanagimoto}\ \emph {et~al.}(2019)\citenamefont
  {Yanagimoto}, \citenamefont {Ng}, \citenamefont {Onodera},\ and\
  \citenamefont {Mabuchi}}]{yanagimoto2019adiabatic}%
  \BibitemOpen
  \bibfield  {author} {\bibinfo {author} {\bibfnamefont {R.}~\bibnamefont
  {Yanagimoto}}, \bibinfo {author} {\bibfnamefont {E.}~\bibnamefont {Ng}},
  \bibinfo {author} {\bibfnamefont {T.}~\bibnamefont {Onodera}},\ and\ \bibinfo
  {author} {\bibfnamefont {H.}~\bibnamefont {Mabuchi}},\ }\bibfield  {title}
  {\bibinfo {title} {Adiabatic fock-state-generation scheme using kerr
  nonlinearity},\ }\href@noop {} {\bibfield  {journal} {\bibinfo  {journal}
  {Physical Review A}\ }\textbf {\bibinfo {volume} {100}},\ \bibinfo {pages}
  {033822} (\bibinfo {year} {2019})}\BibitemShut {NoStop}%
\bibitem [{\citenamefont {Ben~Hayun}\ \emph {et~al.}(2021)\citenamefont
  {Ben~Hayun}, \citenamefont {Reinhardt}, \citenamefont {Nemirovsky},
  \citenamefont {Karnieli}, \citenamefont {Rivera},\ and\ \citenamefont
  {Kaminer}}]{ben2021shaping}%
  \BibitemOpen
  \bibfield  {author} {\bibinfo {author} {\bibfnamefont {A.}~\bibnamefont
  {Ben~Hayun}}, \bibinfo {author} {\bibfnamefont {O.}~\bibnamefont
  {Reinhardt}}, \bibinfo {author} {\bibfnamefont {J.}~\bibnamefont
  {Nemirovsky}}, \bibinfo {author} {\bibfnamefont {A.}~\bibnamefont
  {Karnieli}}, \bibinfo {author} {\bibfnamefont {N.}~\bibnamefont {Rivera}},\
  and\ \bibinfo {author} {\bibfnamefont {I.}~\bibnamefont {Kaminer}},\
  }\bibfield  {title} {\bibinfo {title} {Shaping quantum photonic states using
  free electrons},\ }\href@noop {} {\bibfield  {journal} {\bibinfo  {journal}
  {Science Advances}\ }\textbf {\bibinfo {volume} {7}},\ \bibinfo {pages}
  {eabe4270} (\bibinfo {year} {2021})}\BibitemShut {NoStop}%
\bibitem [{\citenamefont {Lingenfelter}\ \emph {et~al.}(2021)\citenamefont
  {Lingenfelter}, \citenamefont {Roberts},\ and\ \citenamefont
  {Clerk}}]{lingenfelter2021unconditional}%
  \BibitemOpen
  \bibfield  {author} {\bibinfo {author} {\bibfnamefont {A.}~\bibnamefont
  {Lingenfelter}}, \bibinfo {author} {\bibfnamefont {D.}~\bibnamefont
  {Roberts}},\ and\ \bibinfo {author} {\bibfnamefont {A.}~\bibnamefont
  {Clerk}},\ }\bibfield  {title} {\bibinfo {title} {Unconditional fock state
  generation using arbitrarily weak photonic nonlinearities},\ }\href@noop {}
  {\bibfield  {journal} {\bibinfo  {journal} {arXiv preprint arXiv:2103.12041}\
  } (\bibinfo {year} {2021})}\BibitemShut {NoStop}%
\bibitem [{\citenamefont {Rivera}\ \emph {et~al.}(2023)\citenamefont {Rivera},
  \citenamefont {Sloan}, \citenamefont {Salamin}, \citenamefont
  {Joannopoulos},\ and\ \citenamefont
  {Solja{\v{c}}i{\'c}}}]{rivera2023creating}%
  \BibitemOpen
  \bibfield  {author} {\bibinfo {author} {\bibfnamefont {N.}~\bibnamefont
  {Rivera}}, \bibinfo {author} {\bibfnamefont {J.}~\bibnamefont {Sloan}},
  \bibinfo {author} {\bibfnamefont {Y.}~\bibnamefont {Salamin}}, \bibinfo
  {author} {\bibfnamefont {J.~D.}\ \bibnamefont {Joannopoulos}},\ and\ \bibinfo
  {author} {\bibfnamefont {M.}~\bibnamefont {Solja{\v{c}}i{\'c}}},\ }\bibfield
  {title} {\bibinfo {title} {Creating large fock states and massively squeezed
  states in optics using systems with nonlinear bound states in the
  continuum},\ }\href@noop {} {\bibfield  {journal} {\bibinfo  {journal}
  {Proceedings of the National Academy of Sciences}\ }\textbf {\bibinfo
  {volume} {120}},\ \bibinfo {pages} {e2219208120} (\bibinfo {year}
  {2023})}\BibitemShut {NoStop}%
\bibitem [{\citenamefont {Mandel}\ and\ \citenamefont
  {Wolf}(1995)}]{mandel1995optical}%
  \BibitemOpen
  \bibfield  {author} {\bibinfo {author} {\bibfnamefont {L.}~\bibnamefont
  {Mandel}}\ and\ \bibinfo {author} {\bibfnamefont {E.}~\bibnamefont {Wolf}},\
  }\href@noop {} {\emph {\bibinfo {title} {Optical coherence and quantum
  optics}}}\ (\bibinfo  {publisher} {Cambridge university press},\ \bibinfo
  {year} {1995})\BibitemShut {NoStop}%
\bibitem [{\citenamefont {Walls}\ and\ \citenamefont
  {Milburn}(2007)}]{walls2007quantum}%
  \BibitemOpen
  \bibfield  {author} {\bibinfo {author} {\bibfnamefont {D.~F.}\ \bibnamefont
  {Walls}}\ and\ \bibinfo {author} {\bibfnamefont {G.~J.}\ \bibnamefont
  {Milburn}},\ }\href@noop {} {\emph {\bibinfo {title} {Quantum optics}}}\
  (\bibinfo  {publisher} {Springer Science \& Business Media},\ \bibinfo {year}
  {2007})\BibitemShut {NoStop}%
\bibitem [{\citenamefont {Drummond}\ and\ \citenamefont
  {Walls}(1980)}]{drummond1980quantum}%
  \BibitemOpen
  \bibfield  {author} {\bibinfo {author} {\bibfnamefont {P.}~\bibnamefont
  {Drummond}}\ and\ \bibinfo {author} {\bibfnamefont {D.}~\bibnamefont
  {Walls}},\ }\bibfield  {title} {\bibinfo {title} {Quantum theory of optical
  bistability. i. nonlinear polarisability model},\ }\href@noop {} {\bibfield
  {journal} {\bibinfo  {journal} {Journal of Physics A: Mathematical and
  General}\ }\textbf {\bibinfo {volume} {13}},\ \bibinfo {pages} {725}
  (\bibinfo {year} {1980})}\BibitemShut {NoStop}%
\bibitem [{\citenamefont {Siegman}(1986)}]{siegman1986lasers}%
  \BibitemOpen
  \bibfield  {author} {\bibinfo {author} {\bibfnamefont {A.}~\bibnamefont
  {Siegman}},\ }\href@noop {} {\emph {\bibinfo {title} {Lasers}}}\ (\bibinfo
  {publisher} {University Science Books},\ \bibinfo {year} {1986})\BibitemShut
  {NoStop}%
\bibitem [{\citenamefont {Ispasoiu}\ \emph {et~al.}(2002)\citenamefont
  {Ispasoiu}, \citenamefont {Jin}, \citenamefont {Lee}, \citenamefont
  {Papadimitrakopoulos},\ and\ \citenamefont
  {Goodson}}]{ispasoiu_two-photon_2002}%
  \BibitemOpen
  \bibfield  {author} {\bibinfo {author} {\bibfnamefont {R.~G.}\ \bibnamefont
  {Ispasoiu}}, \bibinfo {author} {\bibfnamefont {Y.}~\bibnamefont {Jin}},
  \bibinfo {author} {\bibfnamefont {J.}~\bibnamefont {Lee}}, \bibinfo {author}
  {\bibfnamefont {F.}~\bibnamefont {Papadimitrakopoulos}},\ and\ \bibinfo
  {author} {\bibfnamefont {T.}~\bibnamefont {Goodson}},\ }\bibfield  {title}
  {\bibinfo {title} {Two-photon {Absorption} and {Photon}-number {Squeezing}
  with {CdSe} {Nanocrystals}},\ }\href {https://doi.org/10.1021/nl015636i}
  {\bibfield  {journal} {\bibinfo  {journal} {Nano Letters}\ }\textbf {\bibinfo
  {volume} {2}},\ \bibinfo {pages} {127} (\bibinfo {year} {2002})},\ \bibinfo
  {note} {publisher: American Chemical Society}\BibitemShut {NoStop}%
\end{thebibliography}%
\end{document}


\title{Supplementary Information for: \\Intense squeezed light from lasers with sharply nonlinear gain at optical frequencies}
\author{Linh Nguyen$^{1}$, Jamison Sloan$^{2}$, Nicholas Rivera$^{1,3}$, Marin Solja\v{c}i\'{c}$^{1}$}

\address{$^1$ Department of Physics, Massachusetts Institute of Technology, Cambridge MA, United States. \\
$^2$Department of Electrical Engineering and Computer Science, Massachusetts Institute of Technology, Cambridge MA, United States. \\
$^3$ Department of Physics, Harvard University, Cambridge MA, United States.}
\email{linhnk@mit.edu}


\maketitle

In this supplementary document, we develop the quantum theory of lasers which leads to the generation of macroscopic sub-Poissonian states of light. First, we present the semi-classical rate equations describing the evolution of the polarization, inversion, and photon number. We will consider the case of fast-decaying polarization (class $A$ and $B$ lasers). In that limit, the polarization can be adiabatically eliminated, and the equations of motion for the atomic inversion and the light intensity can be derived. The semi-classical approach can give information about the steady-state light intensity and the transient dynamics (e.g., relaxation oscillations). Then we ``upgrade'' the equations to Heisenberg-Langevin equations through the addition of Langevin forces, enabling us to derive the amplitude noise spectrum and the overall photon number uncertainty.

\subsection{Semiclassical theory of a ``Kerr" laser and adiabatic elimination of class B laser.}

A system of semi-classical theory rate equations for the polarization $\alpha$, atomic inversion $S$, and electromagnetic field $b$ for single mode radiation in a gain medium are given by: 

\begin{align}
    \Dot{b} = (-i\omega_c\left(|b|^2\right) - \frac{\kappa}{2}) b -ig\alpha \tag{S1}\label{eq:S1}\\
    \Dot{\alpha} = (-i\omega_0 - \gamma_\perp) \alpha + igbS \tag{S2}\label{eq:S2}\\
    \Dot{S} = \gamma_\parallel(S_0 - S) + i (g\alpha b^* - g \alpha^* b) \tag{S3}\label{eq:S3}
\end{align}

Here, the only difference with previous treatments of semiclassical laser dynamics (see e.g., \cite{haken1985laser}, \cite{scully1999quantum}) is the intensity-dependent cavity resonant frequency $\omega_c\left(|b|^2\right)$ due to the presence of the Kerr medium. Eq. \ref{eq:S2} admits the formal solution: 
\begin{align}
    \alpha(t) &= \alpha(0) e^{-(i\omega_0 + \gamma_\perp) t} \nonumber \\
    &+ ig \int_0^t \mathrm{d}t' e^{-(i\omega_0 + \gamma_\perp) (t-t')} b(t') S(t')\tag{S4}\label{eq:S4}
\end{align}
where $\gamma S_0 = \Lambda$. $S$ evolves on time scale $\gamma_\parallel^{-1}$; $B$, the slow envelope of $b$ ($b(t) = B(t) e^{-i \omega_c \left(|b|^2\right) t}$), evolves on time scale $\kappa^{-1}$. Therefore, $S$ and $B$ change very slowly over $\gamma_\perp^{-1}$. The polarization can be approximated by:

\begin{align}
    \alpha(t) &\approx \alpha(0) e^{-(i\omega_0 + \gamma_\perp) t} \nonumber\\
    &+ i g B(t) S(t) \int_0^t \mathrm{d}t' e^{-(i\omega_0 + \gamma_\perp) t} e^{i(\omega_0 - \omega_c(|B|^2))t' + \gamma_\perp t'} \nonumber\\
    &\approx \frac{igbS}{i(\omega_0 - \omega_c(|B|^2) + \gamma_\perp}\tag{S5}\label{eq:S5} 
\end{align}
where we have neglected term with $e^{-\gamma_\perp t}$, because we are interested in laser dynamics on timescale $\mathcal{O}(\kappa^{-1}) \gg \gamma_\perp^{-1}$. We also define $\Delta(|B|^2) \equiv \omega_0 - \omega_c(|B|^2)$. Plugging the result from \ref{eq:S5} into \ref{eq:S1} and \ref{eq:S3}, the rate equations reduce to: 
\begin{align}
    \Dot{b} = -i \omega_c(|B|^2) b - \frac{\kappa}{2} b +\frac{g^2 b S}{i\Delta(|B|^2) + \gamma_\perp} \tag{S6}\label{eq:S6}\\
    \Dot{S} = \gamma_\parallel(S_0 - S) - \frac{2 g^2 |B|^2 S}{\gamma_\perp^2 + \Delta^2(|B|^2)} \tag{S7}\label{eq:S7}
\end{align}
Eq. \ref{eq:S7} is identical to Eq. 1b in the main text without the Langevin force. By taking the complex conjugate of Eq. \ref{eq:S6}, we can obtain the differential equation for the photon number $n(t) = b(t)b^{*}(t)$, which is idential to Eq. 1a in the main text without the Langevin forces. 

To derive the steady-state intensity, we set $\Dot{n}$ and $\Dot{S}$ to be zero. The steady-state light intensity satisfies the quadratic equation: 
\begin{equation}
    \beta^2 n_s^2 + n_s\left(- 2\Delta_0 \beta + \frac{2 g^2 \gamma_\perp}{\gamma_\parallel}\right) + \gamma_\perp^2 + \Delta_0^2 - \frac{2 g^2 \gamma_\perp \Lambda}{\kappa \gamma_\parallel} = 0 \tag{S8}\label{eq:S8}
\end{equation}

Eq. \ref{eq:S8} has one positive solution if $\Lambda > \Lambdath(1 + \Delta_0^2/\gamma_\perp^2)$; two positive solutions if $ - 2\Delta_0 \beta + 2 g^2 \gamma_\perp/\gamma_\parallel < 0$ and $\Lambdath\left[1 - g^4/\left(\beta^2 \gamma_\perp^2\right) + 2\Delta_0 g^2/\left(\beta \gamma_\parallel \gamma_\perp\right)\right] < \Lambda < \Lambdath(1 + \Delta_0^2/\gamma_\perp^2)$; otherwise there is no solution. In the two solutions case, for the solution to be stable, it must fall within the decreasing gain region because otherwise an initial small deviation from equilibrium will take the laser far from that equilibrium.

We also consider the transient dynamics of the inversion and the photon number. At time $t$, the photon number deviates by $\delta n(t)$ from the mean photon number, and the inversion deviates by $\delta S(t)$ from the mean inversion. In this paper, we consider the case where the fluctuations of the photon number and inversion are small compared to the mean values (mean-field approximation). We can linearize the system of equations 1 (main text) to become Eq. 2 (main text) without the Langevin force. We may write Eq. 2 (main text) in a compact form $\Dot{\delta V} = M \delta V$. By setting $\delta V(t) = V(0) e^{st}$, we can solve for $s$, given by the eigenvalues of $M$: $s = - \Gamma/2 \pm \sqrt{-\Omega^2}$. When the light intensity is in the decreasing gain region, or $n_s > \Delta_0/\beta - g^2 \gamma_\perp/ (\gamma_\parallel \beta^2)$, the decay rate $\Gamma$ is positive. Thus, it is a stable solution, justifying the intuition above regarding operating in the region of negative differential gain.

\subsection{Quantum Langevin theory of photon number fluctuations in a ``Kerr" laser.}

To study the quantum fluctuations of a nonlinear laser, we add quantum Langevin forces to semi-classical rate equations of the inversion and photon number, which results in Eq. 1 (main text) \cite{quantum_noise}. The linear equation for fluctuations can be written as $\Dot{\delta V} = M \delta V + F$. Consider the solution to the linear equation for the quantum fluctuations in the frequency-domain (defining the Fourier components $\delta V(\omega) = \int_{-\infty}^\infty \mathrm{d}t \delta V(t) e^{i \omega t}$, and $F_i(\omega) = \int_{-\infty}^{\infty} \mathrm{d}t e^{i\omega t} F_i(t)$): 
\begin{align}
    \delta V (\omega) &= - [M + i\omega]^{-1} F(\omega) \nonumber\\
    &= K(\omega) F(\omega) \tag{S9}\label{eq:S9}
\end{align}
where, 
\begin{equation}
    K(\omega) = \frac{-1}{\Omega^2 + \frac{\Gamma^2}{4} - \omega^2 - i\omega \Gamma} \begin{pmatrix}
    i\omega - \Gamma_0 & -\Rsp(n_s) n_s \\
    \kappa + \Gamma_0 - \Gamma & i\omega + \Gamma_0 - \Gamma
    \end{pmatrix} \tag{S10}\label{eq:S10}
\end{equation}

Since $\delta n$ and $\delta S$ are real operators, $\delta V(\omega) = \delta V ^{\dagger} (-\omega)$. Therefore, the overall photon number uncertainty has the form $(\Delta n)^2 = \frac{1}{\pi}\int_0^\infty d\omega \braket{\delta n^\dagger (\omega) \delta n(\omega)}$. We also have 
\begin{align}
    \braket{F_i(\omega) F_j(\omega)} &= \int_{-\infty}^{\infty} \int_{-\infty}^{\infty} \mathrm{d}t \mathrm{d}t' e^{i\omega t} e^{i\omega' t'} 2 \braket{D_{ij}} \delta(t - t')\nonumber\\
    &= 2 \braket{D_{ij}} 2\pi \delta(\omega + \omega') \tag{S11}\label{eq:S11}
\end{align}
Using Eq. \ref{eq:S11} and the Markovian correlations, we can derive the noise spectrum as in Eq. 3 (Main text). The Fano factor has the form  $F = \frac{\Delta n^2}{n} = \frac{\kappa}{\Gamma}\left(\frac{\Gamma_0^2}{\Gamma^2/4 + \Omega^2} + 1\right)$. 

In the case of the ``linear laser'', the Fano factor is found from Eqs. \ref{eq:S9}-\ref{eq:S10} to be $\kappa^2 \gamma_\perp/(2 \Lambda g^2)$ for the resonance case and approximately $\kappa^2 \Delta_0^2/(2 \Lambda g^2 \gamma_\perp)$ for the off-resonance case. 

Meanwhile, for the ``resonant Kerr laser" (of Fig. 2 of the main text), we find an approximate form for the Fano factor as:
\begin{equation}
    F \approx -\frac{\Rsp(n_s)}{\Rsp'(n_s) n_s} = \frac{\gamma_\perp^2 + \beta^2 n_s^2}{2 \beta^2 n_s^2} \tag{S12}\label{eq:S12}
\end{equation}
which decreases as the Kerr strength increases. When $\beta \gg \left[\kappa g^2 \gamma_\perp/ (2 \Lambda \gamma_\parallel)\right]^{1/2}$, $(\beta n_s)^2 \to 2 g^2 \gamma_\perp \Lambda/\left(\kappa \gamma_\parallel \right) \gg \gamma_\perp^2$. So $F \to 1/2$. 

Finally, for the off-resonant Kerr laser analyzed in Fig. 3 (main text), we find for the Fano factor:
\begin{equation}
    F = \frac{\gamma_\perp^2 + (- \Delta_0 + \beta n_s)^2}{2\beta n_s |-\Delta_0 + \beta n_s|} \tag{S13}\label{eq:S13}
\end{equation}
In this case, the Fano factor is extremized when $\Delta_0 \Delta^2 - 2 \gamma_\perp^2 |\Delta| - \Delta_0 \gamma_\perp^2 = 0$, or $|\Delta| \approx \gamma_\perp$. The two minimum values are $F = \gamma_\perp/(\Delta_0 \pm \gamma_\perp)$.

\bibliography{sample}